\documentclass[12pt]{article}

\usepackage{a4}

\usepackage{amsmath}

\usepackage{bm}
\usepackage{amsfonts}

\newcommand{\CC}{\mathbb{C}}
\newcommand{\ZZ}{\mathbb{Z}}
\newcommand{\RR}{\mathbb{R}}
\newcommand{\HH}{\mathbb{H}}

\newcommand{\tr}{{\rm tr}}
\newcommand{\wh}{\widehat}
\newcommand{\wt}{\widetilde}
\newcommand{\ol}{\overline}

\newcommand{\rap}[2]
{\setbox1=\hbox{#1}%
\setbox2=\hbox to\wd1{\hss #2\hss}%
\mbox{\rlap{\box1}\box2}}

\newcommand{\sla}[1]{\rap{$#1$}{$\backslash$}}

\begin{document}\begin{titlepage}
\title{\vspace{-2cm}
\begin{flushright}
\normalsize{TIT/HEP-634\\
April 2014}
\end{flushright}
       \vspace{2cm}
Supersymmetric backgrounds from 5d ${\cal N}=1$ supergravity
       \vspace{2cm}}
\author{
Yosuke Imamura\thanks{E-mail: \tt imamura@phys.titech.ac.jp} $^a$
and
Hiroki Matsuno\thanks{E-mail: \tt matsuno@th.phys.titech.ac.jp} $^a$.
\\[30pt]
{\it $^a$Department of Physics, Tokyo Institute of Technology,}\\
{\it Tokyo 152-8551, Japan}
}
\date{}

\maketitle
\thispagestyle{empty}

\vspace{0cm}

\begin{abstract}
\normalsize
We construct curved backgrounds with Euclidean signature
admitting rigid supersymmetry
by using a 5d ${\cal N}=1$ off-shell Poincar\'e supergravity.
We solve the conditions for the background Weyl multiplet and
vector multiplets
that preserve at least one supersymmetry
parameterized by a symplectic Majorana spinor,
and represent the solution in
terms of several independent fields.
We also show that the
partition function does not depends on
the local degrees of freedom
of the background fields.
Namely, as far as we focus on
a single coordinate patch,
we can freely change the independent fields
by combining $Q$-exact deformations and
gauge transformations.
We also discuss realization of
several known
examples of supersymmetric theories
in curved backgrounds by using the supergravity.
\end{abstract}

\end{titlepage}


\section{Introduction}
Recent progress in
non-perturbative aspects of supersymmetric field theories
owes
a great deal to
the construction of rigid supersymmetry on curved backgrounds.
Pestun \cite{Pestun:2007rz} constructed ${\cal N}=2$
supersymmetric theories on ${\bm S}^4$ and computed the partition function
and the expectation values of circular Wilson loops.
The results play a crucial role in the AGT conjecture \cite{Alday:2009aq},
which relates 4d ${\cal N}=2$ theories and 2d conformal field theories.
The partition function of supersymmetric theories on ${\bm S}^3$
\cite{Kapustin:2009kz,Jafferis:2010un,Hama:2010av}
enables us to perform quantitative checks of dualities among 3d theories
and the AdS$_4$/CFT$_3$ correspondence.
Supersymmetric theories on 3d and 4d squashed spheres
\cite{Hama:2011ea,Imamura:2011wg,Hama:2012bg,Nosaka:2013cpa}
and manifolds with other topologies
\cite{Romelsberger:2007ec,Kim:2009wb,Imamura:2011su}
are also constructed, and the exact partition functions for those
provide useful information
about supersymmetric field theories.

Also in 5d, supersymmetric field theories
are constructed on various curved manifolds.
Theories on round and squashed ${\bm S}^5$ are constructed in
\cite{Kallen:2012cs,Hosomichi:2012ek,Kallen:2012va,Kim:2012av,Imamura:2012xg}.
The perturbative
\cite{Kallen:2012cs,Kallen:2012va,Kim:2012av,Lockhart:2012vp,Imamura:2012bm}
and instanton \cite{Kim:2012qf} partition functions on ${\bm S}^5$ are computed,
and $N^3$ behavior of the free energy of the maximally supersymmetric Yang-Mills theory (SYM)
is confirmed \cite{Kallen:2012zn}.
This is a strong evidence of the close relation \cite{Douglas:2010iu,Lambert:2010iw} between
the 5d SYM and the 6d $(2,0)$ theory realized on a stack of $N$ M5-branes.
Supersymmetric theories on ${\bm S}^4\times{\bm S}^1$
are constructed in \cite{Kim:2012gu,Terashima:2012ra},
and the partition function \cite{Kim:2012gu}
(superconformal index)
provides
an evidence of the existence of non-trivial fixed points with
enhanced global symmetries \cite{Seiberg:1996bd}.
Theories on ${\bm S}^3\times\Sigma$, the product of
three-sphere and a Riemann surface $\Sigma$,
are constructed in \cite{Kawano:2012up,Fukuda:2012jr},
and used to study a conjectured relation
between the 6d $(2,0)$ theory and a $q$-deformed 2d Yang-Mills theory.
Supersymmetric theories on ${\bm S}^2\times M_3$
\cite{Yagi:2013fda,Lee:2013ida,Cordova:2013cea}
are used to confirm predictions of
the 3d/3d correspondence \cite{Dimofte:2011ju}.

A systematic construction of rigid supersymmetric field theories
on curved backgrounds was started in \cite{Festuccia:2011ws}.\footnote
{
See also \cite{Adams:2011vw} for construction of supersymmetric theories
on $AdS_4$ by taking the decoupling limit of supergravity.
}
To obtain rigid supersymmetry on a curved manifold,
we couple matter fields to a background off-shell supergravity multiplet,
and require the supersymmetry transformation
of the gravitino, $\delta_Q\psi_\mu$, to vanish.
If the gravity multiplet contains other fermions their
supersymmetry transformation should also vanish.
By solving these conditions, we obtain
backgrounds that admit rigid supersymmetry.
In 4d, the analysis by using the new minimal supergravity \cite{Sohnius1981}
shows that we can realize at least one rigid supersymmetry
on backgrounds with Hermitian metrics \cite{Klare:2012gn,Dumitrescu:2012ha}.
With the old minimal supergravity \cite{Stelle1978,Ferrara1978}
we can realize a supersymmetry in (squashed) ${\bm S}^4$ or
backgrounds with Hermitian metrics \cite{Dumitrescu:2012at}.
(See also \cite{Jia:2011hw,Samtleben:2012gy} for studies of
4d supersymmetric theories on curved background with the help of supergravity.)
The analysis in \cite{Klare:2012gn,Closset:2012ru} using the
3d version of the new minimal supergravity
shows that the manifold is required to have
the almost contact metric structure
which satisfies a certain integrability condition.
The existence of two or more supersymmetries imposes stronger restrictions.
See \cite{Klare:2012gn,Cassani:2012ri} for analysis from the holographic viewpoint.

In this paper, we realize rigid supersymmetry
on a 5d manifold ${\cal M}$ with Euclidean signature
by using the 5d ${\cal N}=1$ off-shell Poincar\'e supergravity
\cite{Zucker:1999ej,Kugo:2000hn,Kugo:2000af}
whose Weyl multiplet has $40+40$ degrees of freedom \cite{Howe1981}.
The first step of the analysis
with this supergravity
is taken in \cite{Pan:2013uoa},
where the condition associated with the gravitino,
$\delta_Q\psi_\mu=0$, is focused on.
(See also \cite{Gauntlett:2003fk} for supersymmetric backgrounds
in the minimal gauged supergravity
without auxiliary fields \cite{Gunaydin1984},
and \cite{Kehagias:2012fh} for an analysis with
off-shell conformal supergravity with a smaller Weyl multiplet
with $32+32$ degrees of freedom \cite{Gunaydin1985,Bergshoeff:2001hc,Fujita:2001kv,Bergshoeff:2004kh}.)
There is actually another fermion, which we denote by $\eta$,
in the Weyl multiplet,
and thus we should also consider the condition $\delta_Q\eta=0$.
One of the purposes of this paper is to complete this analysis
and to give the solution to the supersymmetry conditions.

Another purpose of this paper is to study supersymmetry-preserving
deformations of the background.
It is often happens that
the partition functions for different backgrounds
are the same.
For example, in the case of ${\bm S}^3$ partition function,
a certain squashed ${\bm S}^3$ gives the same partition function
as the round ${\bm S}^3$ \cite{Hama:2011ea}.
The partition function of another
squashed ${\bm S}^3$ \cite{Imamura:2011wg}
is the same as that for an ellipsoid \cite{Hama:2011ea}.
These facts suggests that the partition function
depends only on a small part of the data of the background.
This is confirmed in
\cite{Alday:2013lba}
by showing that the partition function of a supersymmetric theory
on manifolds with ${\bm S}^3$ topology depends on the background
manifolds only through a single parameter.
Furthermore, \cite{Closset:2013vra} shows
for 3d and 4d cases
that
although supersymmetric backgrounds have functional degrees of freedom
almost all deformations of the background
correspond to $Q$-exact deformations of the action,
and do not affect the partition function.
We perform similar analysis in 5d.

This paper is organized as follows.
In the next section
we solve the conditions $\delta_Q\psi=\delta_Q\eta=0$
and derive the restrictions for the background fields
under the assumption of the existence of at least one
rigid supersymmetry parameterized by a symplectic Majorana spinor.
In Section \ref{def.sec},
we show that all supersymmetry-preserving deformations of the background fields
can be realized by $Q$-exact deformations
and gauge transformations
as far as we focus on a
single coordinate patch.
We also study supersymmetric backgrounds of vector multiplets
in Section \ref{vector.sec}.
In Section \ref{examples.sec} we
discuss realization of
some known examples of supersymmetric theories in curved manifolds
by using the supergravity.
Section \ref{disc.sec} is devoted to discussion.
Notation and conventions are summarized in Appendix.

\section{Supersymmetric backgrounds}\label{susybkg.sec}
\subsection{5d ${\cal N}=1$ off-shell supergravity}\label{sugra.ssec}
The 5d ${\cal N}=1$
off-shell supergravity constructed in \cite{Zucker:1999ej,Kugo:2000hn,Kugo:2000af}
has the following local bosonic symmetries.
\begin{itemize}
\item
The general coordinate invariance
\item
$Sp(2)_L$: The local Lorentz symmetry
\item
$Sp(1)_R$: The local $R$-symmetry
\item
$U(1)_Z$: The gauge symmetry associated with the central charge
\end{itemize}
In addition to these, the formulation in \cite{Kugo:2000hn,Kugo:2000af}
has the local dilatation symmetry.
The corresponding gauge field $b_\mu=\alpha^{-1}\partial_\mu\alpha$ is
pure-gauge, and in this paper we fix the gauge by the condition $b_\mu=0$.

The Weyl multiplet consists of the fields shown in Table \ref{table:weyl}.
In particular, it contains two fermions: $\psi_\mu$ and $\eta$.
\begin{table}[htb]
\caption{Component fields in the Weyl multiplet.
The last two columns show the relation to Zucker's \cite{Zucker:1999ej} and
Kugo-Ohashi's \cite{Kugo:2000hn}
conventions.
We also show the relations among
supersymmetry parameters and fermion bilinears
in the
three conventions in the last two lines.}
\label{table:weyl}
\begin{center}
\begin{tabular}{rlrcccc}
\hline
\hline
& fields & dof & $Sp(1)_R$ & ours & Zucker & KO \\
\hline
bosons & vielbein & $10$ & ${\bm1}$ & $e_\mu^{\wh\nu}$ & $e_\mu^{\wh\nu}$ & $e_\mu^{\wh\nu}$ \\
& $U(1)_Z$ gauge field & $4$ & ${\bm1}$ & $a_\mu$ & $\frac{\kappa}{\sqrt3}A_\mu$ & $-\frac{1}{2\alpha}A_\mu$ \\
& anti-sym. tensor & $10$ & ${\bm1}$ & $v^{\mu\nu}$ & $2\kappa v^{\mu\nu}$ & $2v^{\mu\nu}$ \\
& $Sp(1)_R$ triplet scalars & $3$ & ${\bm3}$ & $t_a$ & $-2i\kappa t_a$ & $t_a$ \\
& $Sp(1)_R$ gauge field & $12$ & ${\bm3}$ & $V_\mu^a$ & $\frac{\kappa i}{2}V_\mu^a$ & $-V_\mu^a$ \\
& scalar & $1$ & ${\bm1}$ & $C$ & $16\kappa C$ & $-4C$ \\
\hline
fermions & gravitino & $32$ & ${\bm2}$ & $\psi_{I\mu\alpha}$ & $\frac{\kappa}{\sqrt2}\psi_{I\mu\alpha}$ & $\psi_{I\mu\alpha}$ \\
& fermion & $8$ & ${\bm2}$ & $\eta_{I\alpha}$ & $8\sqrt2\kappa\lambda_{I\alpha}$ & $-8\wt\chi_{I\alpha}$ \\
\hline
& supersymmetry parameter & $8$ & ${\bm2}$ & $\xi$ & $\frac{1}{\sqrt2}\varepsilon$ & $\varepsilon$ \\
\hline
& fermion bilinears &&& $(\psi\chi)$ & $i(\ol\psi\chi)$ & $i(\ol\psi\chi)$ \\
\hline
\end{tabular}
\end{center}
\end{table}
A supersymmetric background is defined as
a configuration of the Weyl multiplet
that is invariant under the supersymmetry transformation
with a non-vanishing parameter $\xi$.
If we assume $\psi_\mu=\eta=0$ in the background
the transformations of the bosonic
components automatically vanish,
and the nontrivial conditions are $\delta_Q\psi_\mu=\delta_Q\eta=0$.
The transformation laws of the fermions are \cite{Zucker:1999ej,Kugo:2000hn}
\begin{align}
\delta_Q(\xi)\psi_\mu
&=D_\mu\xi
-f_{\mu\nu}\gamma^\nu\xi
+\frac{1}{4}\gamma_{\mu\rho\sigma}v^{\rho\sigma}\xi
-t\gamma_\mu\xi,\nonumber\\
\delta_Q(\xi)\eta&=
-2\gamma_\nu\xi D_\mu v^{\mu\nu}
+\xi C
+4(\sla D t)\xi
+8(\sla f-\sla v)t\xi+\gamma^{\mu\nu\rho\sigma}\xi f_{\mu\nu}f_{\rho\sigma}.
\label{deltapsieta}
\end{align}
See Appendix for the notation of $Sp(1)_R$ and spinor indices.
We treat the transformation parameter $\xi$ as a Grassmann-even variable.
$f_{\mu\nu}=\partial_\mu a_\nu-\partial_\nu a_\mu$ is the $U(1)_Z$ field strength and $D_\mu$ is
the covariant derivative defined by
\begin{equation}
D_\mu=\partial_\mu
+\delta_M(\omega_{\mu \wh\rho\wh\sigma})
-\delta_U(V_\mu^a)
-\delta_Z(a_\mu),
\end{equation}
where $\delta_M$, $\delta_U$, and $\delta_Z$ are
$Sp(2)_L$, $Sp(1)_R$, and $U(1)_Z$ transformations, respectively.
The explicit form of $D_\mu\xi$ is
\begin{equation}
D_\mu\xi
=\partial_\mu\xi+\frac{1}{4}\omega_{\mu\wh\rho\wh\sigma}\gamma^{\wh\rho\wh\sigma}\xi
-V_\mu\xi.
\end{equation}
In this paper terms in transformation laws (Lagrangians)
including two (three) or more fermions are always omitted.

\subsection{Spinor bilinears and orthonormal frame}
In a 5d spacetime with Lorentzian signature,
the parameter $\xi$ of the local ${\cal N}=1$ supersymmetry transformation
is a symplectic Majorana spinor.
Although
we can impose the symplectic Majorana condition
on $\xi$
also in the Euclidean space,
the condition is not the same as that for Lorentzian signature,
and we do not have to impose it.
To study most general case it is desirable to
consider complex spinor without symplectic Majorana condition imposed.
In this paper, however, we restrict ourselves to the case with
$\xi$ satisfying the symplectic Majorana condition.
This is just for simplicity of the analysis.

Following a standard strategy, we define the bilinears of the
spinor $\xi$:
\begin{align}
S=(\xi\xi),\quad
R^\mu=(\xi\gamma^\mu\xi),\quad
J_{\mu\nu}^a=\frac{1}{S}(\xi\tau_a\gamma_{\mu\nu}\xi).
\label{bilinears}
\end{align}
By a Fierz's identity, we can show
\begin{equation}
\gamma_\mu\xi R^\mu=\xi S.
\label{xichirality}
\end{equation}
The following equations are easily derived from this:
\begin{align}
R_\mu R^\mu=S^2,\quad
J_{\mu\nu}^a R^\nu=0,\quad
-\frac{1}{2}\epsilon_{\mu\nu}{}^{\lambda\rho\sigma}R_\lambda J_{\rho\sigma}^a
=SJ_{\mu\nu}^a.
\label{rrjrrj}
\end{align}
Because $\xi$ is a solution to the
first order differential equation
$\delta_Q\psi_\mu=0$,
it is nowhere vanishing and so are the bilinears.
In particular, $S>0$ everywhere.
We assume the vielbein $e_\mu^{\wh\nu}$ is real,
and then $R^\mu$ is real, too.
The existence of the non-vanishing real vector field $R^\mu$ enables us to
treat the background manifold ${\cal M}$ as a fibration over a base manifold
${\cal B}$, at least locally.
In this paper we will not discuss global issues and focus only
on a single coordinate patch.
Let us define the fifth coordinate $x^5$ by
\begin{equation}
R^\mu\partial_\mu=\partial_5,
\end{equation}
and use a local frame with
\begin{equation}
e^{\wh m}=e_n^{\wh m}dx^n,\quad
e^{\wh 5}=S(dx^5+{\cal V}_mdx^m).
\label{localframe}
\end{equation}
With this frame $R^\mu$ has the local components
\begin{equation}
R_{\wh 5}=S,\quad
R_{\wh m}=0.
\end{equation}
The second and third equations in (\ref{rrjrrj}) can be rewritten as
\begin{equation}
J_{\wh m\wh 5}^a=0,\quad
-\frac{1}{2}\epsilon^{(4)}_{\wh m\wh n\wh k\wh l}J_{\wh k\wh l}^a=J_{\wh m\wh n}^a,
\label{rrjrrj2}
\end{equation}
where $\epsilon_{\wh m\wh n\wh k\wh l}^{(4)}=\epsilon_{\wh m\wh n\wh k\wh l\wh 5}$.
The equation (\ref{xichirality}) means that
$\xi$ has positive chirality with respect to $\gamma_{\wh 5}=S^{-1}R^\mu\gamma_\mu$;
\begin{equation}
\gamma_{\wh 5}\xi=\xi.
\end{equation}

A symplectic Majorana spinor $\chi$ belongs to the $({\bm 4},{\bm 2})$ representation
of $Sp(2)_L\times Sp(1)_R$.
Because $Sp(k)=U(k,\HH)$,
we can treat $\chi$ as
a vector with two quaternionic components.
If we use the matrix representation of quaternions
we can represent $\chi$ as
a $4\times 2$ matrix in the form
\begin{equation}
\chi=(\chi_\alpha{}^I)
=\begin{pmatrix}
U \\
D
\end{pmatrix},
\quad
U=U_0{\bm1}_2+iU_a\tau_a,\quad
D=D_0{\bm1}_2+iD_a\tau_a,\quad
U_i,D_i\in\RR.
\label{chimatrix}
\end{equation}
The vector $R_{\wh m}$ breaks the local Lorentz symmetry $Sp(2)_L$
to its subgroup $Sp(1)_l\times Sp(1)_r$, where
$Sp(1)_l$ and $Sp(1)_r$
act on the upper and lower blocks of the 
matrix (\ref{chimatrix}), respectively.

The chirality condition (\ref{xichirality}) implies that
the spinor $\xi$ has the upper block only.
Furthermore, we can choose a gauge such that $U\propto{\bm1}_2$,
and then $\xi$ is given by
\begin{equation}
\xi=(\xi_\alpha{}^I)
=\sqrt{\frac{S}{2}}\begin{pmatrix}
{\bm1}_2 \\
0
\end{pmatrix},
\label{framexi}
\end{equation}
where the normalization is fixed by $S=(\xi\xi)$.
This gauge choice breaks
$Sp(1)_l\times Sp(1)_R$ into its
diagonal subgroup $Sp(1)_D$.
It is obvious in this frame that
the following eight spinors form a basis of the space of symplectic spinors:
\begin{equation}
\xi_\alpha{}^I,\quad
(\gamma_{\wh m})_\alpha{}^\beta\xi_\beta{}^I,\quad
\xi_\alpha{}^J(\tau_a)_J{}^I.
\label{spinbasis}
\end{equation}
An arbitrary spinor can be expanded by
this basis.
For example, $\gamma_{\wh m\wh n}\xi$
is related to $\xi\tau_a$ by
\begin{equation}
\gamma_{\wh m\wh n}\xi
=-\xi\tau_a J_{\wh m\wh n}^a,\quad
\xi\tau_a=\frac{1}{4}J_{\wh m\wh n}^a\gamma^{\wh m\wh n}\xi.
\label{xitaugammaxi}
\end{equation}
The second relation in 
(\ref{xitaugammaxi}) implies that
the three matrices $J^a$ satisfy the same algebra with the Pauli matrices $\tau_a$;
\begin{equation}
J_{\wh m\wh k}^a J_{\wh k\wh n}^b
=\delta_{ab}\delta_{\wh m\wh n}+i\epsilon_{abc}J_{\wh m\wh n}^c.
\end{equation}
Namely, $J^a$ enjoy the quaternion algebra.\footnote{The definition of
$J^a$ differs from the usual definition of the quaternion basis by factor $i$.}

\subsection{$\delta_Q\psi_\mu=0$}
Let us first solve the condition $\delta_Q\psi_\mu=0$,
which is also investigated in \cite{Pan:2013uoa}.
Using the basis
$(\xi,\gamma_{\wh m}\xi,\tau_a\xi)
=(\gamma_{\wh\mu}\xi,\tau_a\xi)$ in
(\ref{spinbasis}) we decompose $\delta_Q\psi_\mu=0$
into the following conditions:
\begin{align}
0=(\xi\gamma_{\wh\lambda}\delta_Q\psi_{\wh\mu})
&=\frac{1}{2}D_{\wh\mu}R_{\wh\lambda}
-Sf_{\wh\mu\wh\lambda}
-\frac{S}{4}\epsilon_{\wh 5\wh\mu\wh\lambda\wh\rho\wh\sigma}v^{\wh\rho\wh\sigma}
+St_aJ_{\wh\mu\wh\lambda}^a,\label{kv}\\
0=(\xi\tau_a\delta_Q\psi_{\wh\mu})
&=(\xi\tau_aD_{\wh\mu}\xi)
+\frac{1}{4}(\xi\tau_a\gamma_{\wh\mu\wh\rho\wh\sigma}\xi)v^{\wh\rho\wh\sigma}
-R_{\wh\mu} t_a.
\label{xitaidelpsi}
\end{align}
The symmetric part of (\ref{kv}),
$D_{\{\wh\mu}R_{\wh\lambda\}}=0$,
means that $R^\mu$ is a Killing vector.
We can take an $Sp(1)_D\times Sp(1)_r$ gauge such that
\begin{equation}
\partial_5 e_n^{\wh m}=\partial_5S=\partial_5{\cal V}_m=0,
\end{equation}
and then $e_n^{\wh m}$, $S$, and ${\cal V}_m$ can be treated as
fields on the base manifold ${\cal B}$.
The $(\wh\lambda,\wh\mu)=(5,m)$ components of 
(\ref{kv}) give
\begin{equation}
f_{m5}=\frac{1}{2}\partial_mS.
\label{fm5eq}
\end{equation}
From the integrability condition $\partial_nf_{m5}=\partial_mf_{n5}$
and the Bianchi identity for $f_{\mu\nu}$ we obtain
$\partial_5f_{mn}=0$.
This means that the $U(1)_Z$ gauge field $a_\mu$ is essentially
a gauge field on ${\cal B}$.
(\ref{fm5eq})
can be solved, up to $U(1)_Z$ gauge transformation,
by
\begin{equation}
a=a_mdx^m+\frac{1}{2}Sdx^5,\quad
\partial_5a_m=0.
\end{equation}

For later use we give the non-vanishing components of the spin connection.
\begin{equation}
\omega_{\wh k-\wh m\wh n}
=\omega_{\wh k-\wh m\wh n}^{(4)},\quad
\omega_{\wh m-\wh n\wh 5}
=\omega_{\wh 5-\wh n\wh m}
=\frac{S}{2}{\cal W}_{\wh m\wh n}
=\frac{1}{S}D_{\wh m}R_{\wh n}
,\quad
\omega_{\wh 5-\wh 5\wh m}=\frac{1}{S}\partial_{\wh m}S=2f_{\wh m\wh 5}.
\end{equation}
$\omega_{\wh k-\wh m\wh n}^{(4)}$ is the spin connection
in the base manifold ${\cal B}$ defined with the vielbein $e_m^{\wh n}$ and
${\cal W}_{mn}=\partial_m{\cal V}_n-\partial_n{\cal V}_m$
is the field strength of ${\cal V}_m$.

The anti-symmetric part of 
(\ref{kv}) can be used to represent the horizontal part of
$v^{\mu\nu}$ in terms of other fields;
\begin{equation}
v_{\wh p\wh q}
=
\epsilon_{\wh p\wh q\wh m\wh n}^{(4)}
\left(\frac{S}{4}{\cal W}_{\wh m\wh n}
-f_{\wh m\wh n}
+t_aJ_{\wh m\wh n}^a
\right).
\end{equation}

By using (\ref{framexi}) we obtain
\begin{equation}
(\xi\tau_a D_\mu\xi)
=
\frac{S}{4}\omega_{\mu\wh p\wh q}J^a_{\wh p\wh q}
-V_\mu^a S,
\end{equation}
and we can solve
(\ref{xitaidelpsi}) with respect to $V_\mu^a$
and obtain
\begin{equation}
V_{\wh 5}^a
=\frac{1}{4}\omega_{\wh 5\wh p\wh q}J^a_{\wh p\wh q}
+\frac{1}{4}J_{\wh p\wh q}^av^{\wh p\wh q}
-t_a,\quad
V_{\wh m}^a
=\frac{1}{4}\omega_{\wh m\wh p\wh q} J^a_{\wh p\wh q}
+\frac{1}{2} J^a_{\wh m\wh p}v^{\wh p\wh 5}.
\label{vsu2is}
\end{equation}
Now we have completely solved $\delta_Q\psi_\mu=0$.
Independent fields
are
\begin{equation}
e_n^{\wh m}(x^m),\quad
S(x^m),\quad
{\cal V}_m(x^m),\quad
a_m(x^m),\quad
v_{\wh m\wh 5}(x^m,x^5),\quad
t_a(x^m,x^5),\quad
C(x^m,x^5),
\end{equation}
and the other fields are represented by these fields.

\subsection{$\delta_Q\eta=0$}
By using the spinor basis (\ref{spinbasis})
we decompose $\delta_Q\eta=0$ into the following equations.
\begin{align}
0&=S^{-1}(\xi\delta_Q\eta)
=
-2D_\mu v^{\mu\wh 5}
+C
+4t_a J_{\wh m\wh n}^a(f^{\wh m\wh n}-v^{\wh m\wh n})
+\epsilon^{(4)}_{\wh m\wh n\wh p\wh q} f^{\wh m\wh n}f^{\wh p\wh q},
\label{ceq}\\
0&=S^{-1}(\xi\gamma_{\wh m}\delta_Q\eta)
=
-2D^\lambda v_{\lambda\wh m}
+4 J^a_{\wh m\wh n} D^{\wh n} t_a
+8t_a J^a_{\wh m\wh p}(f^{\wh p\wh 5}-v^{\wh p\wh 5})
+4\epsilon_{\wh m\wh p\wh q\wh r}^{(4)} f^{\wh p\wh q}f^{\wh r\wh 5},
\label{lasteq}\\
0&=S^{-1}(\xi\tau_a\delta_Q\eta)
=4D_{\wh 5} t_a
+4i\epsilon_{abc} J^c_{\wh m\wh n} t_b(f-v)^{\wh m\wh n}.
\label{d5t}
\end{align}
(\ref{ceq}) is the only condition including $C$,
and can be used to determine $C$.
(\ref{lasteq}) and (\ref{d5t}) are drastically simplified
if we substitute the solution of $\delta_Q\psi_\mu=0$;
\begin{align}
0=S^{-1}(\xi\gamma_{\wh m}\delta_Q\eta)
=2\partial_{\wh 5}v_{\wh m\wh 5},\quad
0=S^{-1}(\xi\tau_a\delta_Q\eta)
=4\partial_{\wh 5} t_a.
\end{align}
Namely, $v_{\wh m\wh 5}$ and $t_a$ are $x^5$ independent.
After all,
we have obtained the following solution:
\begin{align}
\xi_\alpha{}^I&=
\sqrt{\frac{S}{2}}
\begin{pmatrix}
{\bm1}_2\\
0
\end{pmatrix},\nonumber\\
e^{\wh m}_n&=(\mbox{indep.}),\nonumber\\
e_5^{\wh 5}&=S\quad(\mbox{indep.}),\nonumber\\
{\cal V}_m&=(\mbox{indep.}),\nonumber\\
a_{\wh m}&=(\mbox{indep.}),\nonumber\\
a_5&=\frac{1}{2}S,\nonumber\\
v_{\wh p\wh q}
&=
\epsilon_{\wh p\wh q\wh m\wh n}^{(4)}
\left(\frac{S}{4}{\cal W}_{\wh m\wh n}
-f_{\wh m\wh n}
+t_a J_{\wh m\wh n}^a
\right),\nonumber\\
v^{\wh m\wh 5}&=(\mbox{indep.}),\nonumber\\
t_a&=(\mbox{indep.}),\nonumber\\
V_{\wh m}^a
&=\frac{1}{4}\omega^{(4)}_{\wh m\wh p\wh q} J^a_{\wh p\wh q}
+\frac{1}{2} J^a_{\wh m\wh p}v^{\wh p\wh 5},\nonumber\\
V_{\wh 5}^a
&=
\frac{1}{2} J_{\wh m\wh n}^a
\left(f_{\wh m\wh n}-\frac{S}{2}{\cal W}_{\wh m\wh n}\right)
+t_a,\nonumber\\
C&=
2D_{\wh m}^{(4)}v^{\wh m\wh 5}
+4t_a J_{\wh m\wh n}^af_{\wh m\wh n}
+32t_at_a
-\epsilon^{(4)}_{\wh m\wh n\wh p\wh q}
\left(f^{\wh m\wh n}-\frac{S}{2}{\cal W}^{\wh m\wh n}\right)
\left(f^{\wh p\wh q}-\frac{S}{2}{\cal W}^{\wh p\wh q}\right).
\label{gensol}
\end{align}
``(indep.)'' means that the field is an independent field.
All the fields are $x^5$-independent.
This is in fact a direct consequence of the
algebra.
From the commutation relation (2.45) in \cite{Kugo:2000hn},
we obtain
\begin{align}
\delta_Q(\xi)^2&=
R^\mu D_\mu
-\delta_M\left(2Sf_{\wh\mu\wh\nu}
+\frac{1}{2}\epsilon_{\wh\mu\wh\nu\lambda\rho\sigma}R^\lambda v^{\rho\sigma}
-2S J_{\wh\mu\wh\nu}^at_a\right)
\nonumber\\&
+\delta_Z\left(\frac{1}{2}S\right)
+\delta_U\left(-3St_a-\frac{S}{2} J^a_{\wh m\wh n}(f^{\wh m\wh n}-v^{\wh m\wh n})\right)
\nonumber\\&
+(\mbox{terms with $\eta$ or $\psi_\mu$}).
\end{align}
In the background (\ref{gensol}), the right hand side reduces to the $x^5$ derivative;
\begin{equation}
\delta_Q(\xi)^2=
R^\mu D_\mu
-\delta_M(R^\lambda\omega_{\lambda-\wh\mu\wh\nu})
+\delta_Z(R^\mu a_\mu)
+\delta_U(R^\mu V_\mu^a)
=\partial_5.
\end{equation}
Therefore, a $\delta_Q(\xi)$-invariant background
is also invariant
under the isometry $\partial_5$.

\section{$Q$-exact deformations}\label{def.sec}
The solution obtained in the previous section
depends on several functions and has large degrees of freedom.
However, as we will show in this section,
only small part of them can affect the partition function.

Let $S_0$ be the action of a supersymmetric theory on a supersymmetric
background given by
the solution (\ref{gensol}).
A small deformation around the background
induces the change of the action
\begin{equation}
S_1
=
\int d^5x\sqrt{g}\left[
-\delta e_\mu^{\wh\nu}T_{\wh\nu}^\mu
+\delta V_\mu^aR_a^\mu
+(\delta\psi_\mu S^\mu)
-\delta a_\mu J^\mu
+\delta v^{\mu\nu}M_{\mu\nu}
+\delta C\Phi
+(\delta\eta\chi)
+\delta t_aX_a\right],
\label{lprimes3}
\end{equation}
where the set of the operators
\begin{align}
&R_a^\mu(12),\quad
S_{I\alpha}^\mu(32),\quad
T^{\mu\nu}(10),\quad
J^\mu(4),\quad
\nonumber\\&
M^{\mu\nu}(10),\quad
\Phi(1),\quad
\chi_{I\alpha}(8),\quad
X_a(3),
\label{sources}
\end{align}
forms the supercurrent multiplet
with $40+40$ degrees of freedom.
$S^\mu_{I\alpha}$ and $\chi_{I\alpha}$ are fermionic
and the others are bosonic.
The numbers in the parenthesis represent
the degrees of freedom of the operator.

If the change of the background fields are consistent
to the solution (\ref{gensol})
$S_1$ is $Q$-invariant and
the deformed action $S_0+S_1$
gives the supersymmetric theory on the
deformed background.
We would like to consider the problem whether such a supersymmetric
deformation affects the partition function.
If the deformation $S_1$ is $Q$-exact as well as $Q$-invariant,
it does not change the partition function.
A $Q$-exact deformation that is regarded as
a change of the bosonic background fields
in general has the form
\begin{equation}
\delta_Q(\xi)\int\sqrt{g}d^5x\left[
H_\mu S^\mu
+
K\chi\right],
\label{qexactaction}
\end{equation}
where $H_\mu$ and $K$ are
vectorial-spinor and spinor coefficient functions.
Both $H_\mu$ and $K$ are Grassmann-even.
Because $\delta_Q(\xi)^2=\partial_5$
for the action (\ref{qexactaction}) to be $Q$-invariant
the functions $H_\mu$ and $K$ should be $x^5$-independent.

$\delta_QS^\mu$ and $\delta_Q\chi$ are determined as follows.
For an arbitrary deformation that may not preserve the supersymmetry
$S_1$ is invariant under the supersymmetry if we
transform both the Weyl multiplet and matter fields.
The transformation laws of the bosonic components of
the Weyl multiplet are \cite{Zucker:1999ej,Kugo:2000hn} 
\begin{align}
\delta_Q e_\mu^{\wh\nu}
&=-2(\xi\gamma^{\wh\nu}\psi_\mu)
,\nonumber\\
\delta_Q a_\mu
&=-(\xi\psi_\mu)
,\nonumber\\
\delta_Q V_\mu^a
&=-\frac{1}{4}(\xi\tau_a\gamma_\mu\eta)
+(\xi\tau_a\gamma^\lambda R_{\lambda\mu}(Q))
+(\xi\tau_a\gamma^{\rho\sigma}f_{\rho\sigma}\psi_\mu)
-(\xi\tau_a\gamma^{\rho\sigma}v_{\rho\sigma}\psi_\mu)
+6(\xi\psi_\mu)t_a
,\nonumber\\
\delta_Q t_a
&=-\frac{1}{4}(\xi\tau_a\eta)
,\nonumber\\
\delta_Q v_{\wh\mu\wh\nu}
&=\frac{1}{2}(\xi\gamma_{\wh\mu\wh\nu\wh\rho\wh\sigma}R^{\wh\rho\wh\sigma}(Q))
+\frac{1}{2}(\xi\gamma_{\wh\mu\wh\nu}\eta)
,\nonumber\\
\delta_Q C
&=-(\xi\wh{\sla D}\eta)
-11(\xi t\eta)
-\frac{3}{4}(\xi\gamma_{\mu\nu}v^{\mu\nu}\eta)
-4(\xi t\gamma^{\mu\nu}R_{\mu\nu}(Q)),
\label{susyb}
\end{align}
where $R_{\mu\nu}(Q)$ and $\wh D_\mu \eta$ are defined by
\begin{align}
R_{\mu\nu}(Q)&=
2D_{[\mu}\psi_{\nu]}
+\frac{1}{2}\gamma_{\rho\sigma[\mu}\psi_{\nu]}v^{\rho\sigma}
+2\gamma^\rho\psi_{[\mu}f_{\nu]\rho}
-2\gamma_{[\mu}t\psi_{\nu]},\nonumber\\
\wh D_\mu\eta
&=D_\mu\eta-\delta_Q(\psi_\mu)\eta.
\end{align}
By requiring the $Q$-invariance of $S_1$
we can determine the transformation laws of the
fields in (\ref{sources}).
For example, the transformation of $\chi$ is
\begin{equation}
\delta_Q\chi=\frac{1}{4}\tau_a\gamma_\mu\xi R_a^\mu
+\frac{1}{4}\tau_a\xi X_a
-\frac{1}{2}\gamma_{\mu\nu}\xi M^{\mu\nu}
+\gamma^\mu\xi D_\mu\Phi
+f_{\mu\nu}\gamma^{\mu\nu}\xi\Phi
+16t\xi\Phi.
\end{equation}

Let us consider the second term in the $Q$-exact action (\ref{qexactaction}).
It is convenient to expand the spinor function $K$ by the basis in (\ref{spinbasis}) as
\begin{equation}
K=k\xi+\frac{4}{S}k^a\xi\tau_a-\frac{2}{S}k^{\wh m}\xi\gamma_{\wh m}.
\label{fexpand}
\end{equation}
The first term in (\ref{fexpand}), $k\xi$, gives the
action
\begin{equation}
\delta_Q\int d^5x\sqrt{g}k(\xi\chi)
=\int d^5x\sqrt{g}k\partial_5\Phi.
\end{equation}
This is total derivative, and
does not give a non-trivial deformation of the theory.

The second term in (\ref{fexpand}) gives
\begin{align}
&\delta_Q\int d^5x\sqrt{g}\frac{4}{S}k^a(\xi\tau_a\chi)
\nonumber\\
&=\int d^5x\sqrt{g}\left(
k^aR_a^{\wh 5}
+k^aX_a
-2k^a J^a_{\wh m\wh n}M^{\wh m\wh n}
+4k^a J^a_{\wh m\wh n}f^{\wh m\wh n}\Phi
+64k^at_a\Phi
\right).
\label{xitaudchi}
\end{align}
Comparing this to 
(\ref{lprimes3}), we find that the addition of
(\ref{xitaudchi}) to the action is equivalent to the background deformation
\begin{align}
&\delta t_a=k^a,\quad
\delta V_{\wh 5}^a=k^a,\quad
\delta v^{\wh m\wh n}=-2k^a J_{\wh m\wh n}^a,\quad
\delta C=4k^a J_{\wh m\wh n}^af^{\wh m\wh n}
+64k^at_a,\nonumber\\
&
\delta e_\mu^{\wh\nu}=\delta a_\mu=\delta v^{\wh m\wh 5}=\delta V_{\wh m}^a=0.
\label{tdeformation}
\end{align}
These variations are consistent to the
solution
(\ref{gensol}).
We obtain (\ref{tdeformation}) by shifting $t_a$ by
\begin{equation}
t_a\rightarrow t_a+k^a,
\end{equation}
and keeping other independent fields intact.

Similarly, the addition of the $Q$-exact action
\begin{align}
&\delta_Q
\int d^5x\sqrt{g}\left(
-\frac{2}{S}k^{\wh m}(\xi\gamma_{\wh m}\chi)\right)
\nonumber\\
&=
\int d^5x\sqrt{g}\left(
-\frac{1}{2}k^{\wh m}J_{\wh m\wh n}^aR_a^{\wh n}
+2k^{\wh m}M^{\wh m5}
+2(D_{\wh m}^{(4)}k^{\wh m})\Phi\right)
\end{align}
corresponding to the third term in (\ref{fexpand})
is equivalent to the changes of the background fields
\begin{align}
&\delta v^{\wh m\wh 5}=k^{\wh m},\quad
\delta V_{\wh m}^a=\frac{1}{2}J^a_{\wh m\wh n}k^{\wh n},\quad
\delta C=2D_{\wh m}^{(4)}k^{\wh m},\nonumber\\
&\delta e_\mu^{\wh\nu}=\delta a_\mu=\delta v^{\wh m\wh n}=\delta t_a=\delta V_{\wh 5}^a=0.
\label{vm5def}
\end{align}
These variations are again consistent to the solution (\ref{gensol}),
and generated by the shift of the independent field $v^{\wh m\wh 5}$ by
\begin{equation}
v^{\wh m\wh 5}\rightarrow v^{\wh m\wh 5}+k^{\wh m}.
\label{vm5deformation}
\end{equation}

Before considering $Q$-exact terms
made from the supersymmetry current $S^\mu$, which are
expected to be more complicated,
it is convenient to simplify the prescription used above to obtain the
$Q$-exact deformations.
A small deformation of the theory is schematically expressed as
\begin{equation}
S_1=A_i^BJ_i^B+A_i^FJ_i^F,
\end{equation}
where $(A_i^B,A_i^F)$ is a small variation of the Weyl multiplet
around a supersymmetric background and $(J_i^B,J_i^F)$ is
the multiplet of currents.
The superscripts `$B$' and `$F$' indicate
the bosonic and fermionic statistics, respectively.
The index $i$ collectively represents all indices of fields
including the coordinates $x^\mu$.
The transformation laws of the fermionic components $J_i^F$ of the current multiplet
are obtained by requiring the cancellation
\begin{equation}
\delta_QA_i^BJ_i^B-A_i^F\delta_QJ_i^F=0.
\end{equation}
We only need to consider linear order terms with respect to fermions, and
the transformation of bosonic components $A_i^B$ of the Weyl multiplet
can be written as
\begin{equation}
\delta_QA_i^B=A_j^FM_{ji},
\label{dqa}
\end{equation}
where $M_{ji}$ are functions of bosonic fields.
Then the transformation of $J_i^F$ is
\begin{equation}
\delta_QJ_j^F=M_{ji}J_i^B,
\end{equation}
and the general $Q$ exact term can be written as
\begin{equation}
\delta_Q(f_jJ_j^F)=f_jM_{ji}J_i^B,
\end{equation}
where $f_j$ are Grassmann-even deformation parameters.
This can be interpreted as the following deformation of the background.
\begin{equation}
A_i^B=f_jM_{ji}.
\end{equation}
This is nothing but the supersymmetry transformation
(\ref{dqa}) with the fermion fields $A_j^F$ replaced by
the parameters $f_j$.
Namely, changes of the background
which are realized by $Q$-exact deformations
are obtained from the supersymmetry transformation laws
(\ref{susyb}) by replacing fermions by
deformation parameters.
Indeed,
the deformations
(\ref{tdeformation})
and (\ref{vm5def})
are respectively obtained
from (\ref{susyb})
by the replacements
\begin{equation}
(\psi_\mu,\eta)\rightarrow
\left(0,-\frac{4}{S}k^a\tau_a\xi\right),\quad
(\psi_\mu,\eta)\rightarrow
\left(0,-\frac{2}{S}k^{\wh m}\gamma_{\wh m}\xi\right).
\end{equation}

Now let us consider $Q$-exact terms
including $\delta_QS^\mu$
by using this method.
The corresponding background deformation
can be obtained from the transformation laws (\ref{susyb})
by the replacement
\begin{equation}
(\psi_\mu,\eta)\rightarrow(H_\mu,0).
\end{equation}
We expand the function $H_\mu$ by the spinor basis
as
\begin{equation}
H_\mu=-\frac{1}{2S}h_\mu\xi+\frac{1}{S}h_\mu^a\tau_a\xi
-\frac{1}{2S}h_\mu^{\wh m}\gamma_{\wh m}\xi.
\end{equation}
The deformation parameters $h_\mu$, $h_\mu^a$, and $h_\mu^{\wh m}$
are arbitrary functions on the base manifold ${\cal B}$.

The variations of the independent fields
in the deformation by the parameter $h_\mu$ are
\begin{equation}
\delta S=h_5,\quad
\delta {\cal V}_{\wh m}=\frac{1}{S}h_{\wh m},\quad
\delta a_{\wh m}=\delta v_{\wh m\wh 5}=\delta t_a=0.
\end{equation}
The variations of the dependent fields are obtained
from the solution (\ref{gensol}).
By this $Q$-exact deformation we can freely change the functions $S$
and ${\cal V}_{\wh m}$.

The deformation by the parameter $h_\mu^a$ is
\begin{equation}
\delta v_{\wh m\wh 5}
=4i J_{\wh m\wh n}^ah_{\wh n}^bt_c\epsilon_{abc},\quad
\delta S=\delta{\cal V}_m=\delta a_m=\delta t_a=0.
\end{equation}
This is not independent of the deformation
(\ref{vm5deformation}).
Finally, the deformation by the parameter $h_\mu^{\wh m}$ is
\begin{align}
&\delta e_\mu^{\wh m}=h_\mu^{\wh m},\quad
\delta e_\mu^{\wh 5}=
\delta a_\mu=
\delta t_a=0,\nonumber\\
&\delta v_{\wh m\wh 5}
=\epsilon_{\wh m\wh p\wh q\wh k}^{(4)}
\left(\frac{1}{2}D_{\wh p}^{(4)}h_{\wh q}{}^{\wh k}
+\frac{1}{4}{\cal W}_{\wh m\wh n}h_5{}^{\wh k}
-h_{\wh q}{}^{\wh k}v^{\wh p\wh 5}
\right)
+h_{\wh q}{}^{\wh m}v^{\wh q\wh 5}
-h_{\wh q}{}^{\wh q}v^{\wh m\wh 5}.
\end{align}
The change of $v_{\wh m\wh 5}$ can be absorbed by the deformation
(\ref{vm5deformation}),
and we are not interested in it.
By using the parameter $h_n^{\wh m}$
we can freely change the vielbein $e_m^{\wh n}$ of the base manifold ${\cal B}$.
The deformation with the parameter $h_5^{\wh m}$ breaks the choice of the
gauge (\ref{localframe}) for the vielbein.
To recover $e_5^{\wh m}=0$,
we should perform the compensating local Lorentz transformation
\begin{equation}
\delta_M(\lambda_{\wh\mu\wh\nu})e_5^{\wh m}=-h_5^{\wh m},\quad
\lambda_{\wh m\wh 5}=-\frac{1}{S}h_5^{\wh m},\quad
\lambda_{\wh m\wh n}=0.
\end{equation}
This transformation, in turn, changes the vector field $a_{\wh m}$ by
\begin{equation}
\delta_M(\lambda_{\wh\mu\wh\nu})a_{\wh m}=-\frac{1}{2S}h_5^{\wh m}.
\end{equation}
As a result, we can freely change $a_{\wh m}$ by
using the combination of the $Q$-exact deformation
with the parameter $h_5^{\wh m}$ and the compensating
$\delta_M$ transformation.

After all,
by using $Q$-exact deformations and gauge transformations,
we can freely change all the
independent fields.
Of course this does not mean that the partition function does not
depend on the background at all.
To clarify the background dependence of the
partition function,
careful analysis of the
global structure of the background is needed.

\section{Background vector multiplets}\label{vector.sec}
In addition to the Weyl multiplet,
we can introduce vector multiplets
as background fields coupling to global symmetry currents.
A vector multiplet consists of
the component fields shown in Table \ref{table:vector}.
\begin{table}[htb]
\caption{
Component fields in a vector multiplet.
Relation to Kugo-Ohashi's convention is also shown.}
\label{table:vector}
\begin{center}
\begin{tabular}{rlrccc}
\hline
\hline
& fields & dof & $Sp(1)_R$ & ours & KO \\
\hline
bosons & gauge field & $4$ & ${\bm1}$ & $A_\mu$ & $-igW_\mu$ \\
& scalar & $1$ & ${\bm1}$ & $\phi$ & $gM$ \\
& auxiliary fields & $3$ & ${\bm3}$ & $D_a$ & $2gY_a$ \\
fermion & gaugino & $8$ & ${\bm2}$ & $\lambda$ & $-2ig\Omega$ \\
\hline
& prepotential &&& ${\cal F}$ & $-\frac{1}{2}{\cal N}$ \\
\hline
\end{tabular}
\end{center}
\end{table}
The transformation laws for those are
(Eq. (3.2) in \cite{Kugo:2000hn})
\begin{align}
\delta_Q(\xi)\lambda&=
-\sla F\xi
+2i\phi \sla f\xi
+i(\sla D\phi)\xi
+i D\xi,\nonumber\\
\delta_Q(\xi) A_\mu&
=-(\xi\gamma_\mu\lambda)
-2i(\xi\psi_\mu)\phi,\nonumber\\
\delta_Q(\xi)\phi&=i(\xi\lambda),\nonumber\\
\delta_Q(\xi) D_a&=i(\xi\tau_a\wh{\sla D}\lambda)
-i(\xi\tau_a[\phi,\lambda])
-\frac{i}{2}(\xi\tau_a\sla v\lambda)
-i(\xi\tau_a t\lambda)
+4i(\xi\lambda)t_a,
\label{vectortrlaws}
\end{align}
where $F_{\mu\nu}=\partial_\mu A_\nu-\partial_\nu A_\mu-i[A_\mu,A_\nu]$
is the gauge field strength and $\wh D_\mu\lambda$ is
the supercovariant derivative
\begin{equation}
\wh D_\mu\lambda=D_\mu\lambda-\delta_Q(\psi_\mu)\lambda.
\end{equation}

In the presence of the background vector multiplets
we should impose the condition $\delta_Q\lambda=0$.
For simplicity, we consider
a $U(1)$ vector multiplet.
We decompose the condition into
the following two.
\begin{align}
0=(\xi\gamma_\mu\delta_Q\lambda)
&=
-F_{\mu\nu}R^\nu
+2i\phi f_{\mu\nu}R^\nu
+iSD_\mu\phi,\label{ximuql}\\
0=(\xi\tau_a\delta_Q\lambda)
&=-\frac{S}{2}(F^{\mu\nu}-2i\phi f^{\mu\nu}) J^a_{\mu\nu}+iS D_a.
\label{xiaql}
\end{align}
From (\ref{ximuql}) we obtain
\begin{align}
D_5\phi=0,\quad
F_{m5}=iD_m(S\phi).
\label{eq66}
\end{align}
The solution of (\ref{eq66}) together with
$D_a$ represented in terms of other fields by solving (\ref{xiaql})
is
\begin{align}
\phi&=(\mbox{indep.}),\nonumber\\
A_5&=iS\phi,\nonumber\\
A_m&=(\mbox{indep.}),\nonumber\\
D_a&=-\frac{i}{2}(F_{\wh m\wh n}-2i\phi f_{\wh m\wh n}) J^a_{\wh m\wh n},
\label{vectsol}
\end{align}
up to gauge transformation.

Next, let us specify the degrees of freedom realized by
$Q$-exact deformations.
As explained in the previous section,
such deformations can be easily obtained from
the transformation laws of the bosonic components in
(\ref{vectortrlaws})
by replacing the fermion $\lambda$ by deformation parameters.
The replacement
$\lambda\rightarrow -S^{-1}f_\mu \gamma^\mu\xi$
gives
\begin{equation}
\delta A_\mu=f_\mu,\quad
\delta \phi=-iS^{-1}f_5,
\label{eq68}
\end{equation}
while $\lambda\rightarrow f^a\tau_a\xi$ does not give
non-trivial deformation.
By using (\ref{eq68})
we can freely change the independent fields
in (\ref{vectsol}), at least locally.

\section{Examples}\label{examples.sec}
\subsection{Conformally flat backgrounds}\label{conf.ssec}
For a given superconformal field theory on the flat background,
it is straightforward to obtain the theory in a conformally flat
background by a Weyl transformation that maps the
flat space to the conformally flat background.
The parameter $\xi$ of the superconformal transformation
on the background
satisfies
the Killing spinor equation
\begin{equation}
D_\mu\xi=\gamma_\mu\kappa\quad
\exists\kappa.
\label{generalkilling}
\end{equation}

For vector multiplets, the Weyl transformation gives the
superconformal transformation laws
\begin{align}
\delta_{\rm SC} A^i_\mu&=-(\xi\gamma_\mu\lambda^i),
\nonumber\\
\delta_{\rm SC}\phi^i&=i(\xi\lambda^i),
\nonumber\\
\delta_{\rm SC}\lambda^i
&=-\sla F^i\xi
+i(\sla D\phi^i)\xi
+iD'^i\xi
+2i\kappa\phi^i
,\nonumber\\
\delta_{\rm SC} D'^i_a
&=i(\xi\tau_a\gamma^\mu D_\mu\lambda^i)
-i(\xi\tau_a[\phi,\lambda]^i)
-i(\kappa\tau_a\lambda^i),
\label{deltavector}
\end{align}
where $D'$ is the auxiliary field,
whose definition is different from the previous auxiliary field $D$.
The relation between $D$ and $D'$ will be shown later.
We use $i,j,\ldots$ for adjoint indices of the gauge group.
The superconformal Lagrangian is
\begin{equation}
e^{-1}{\cal L}_{\rm SC}^{(V)}
=e^{-1}{\cal L}_0^{(V)}|_{\rm conf}+\frac{R}{4}{\cal F},
\label{wel}
\end{equation}
where ${\cal L}_0^{(V)}$ is the superconformal Lagrangian on the flat background
covariantized with respect to
the local symmetries listed in \ref{sugra.ssec}:
\begin{align}
e^{-1}{\cal L}_0^{(V)}
&=
-\frac{1}{2}{\cal F}_i[\lambda,\lambda]^i
\nonumber\\
&+{\cal F}_{ij}\left(
\frac{1}{4}F_{\mu\nu}^iF^{\mu\nu j}
+\frac{1}{2}D_\mu\phi^i D^\mu\phi^j
-\frac{1}{2}D'^i_aD'^j_a
-\frac{1}{2}\lambda^i\sla D\lambda^j
\right)
\nonumber\\
&+{\cal F}_{ijk}
\left(
\frac{i}{6}[\mbox{CS}]_5^{ijk}
+\frac{1}{4}\lambda^i(i\sla F^j+D'^j)\lambda^k
\right).
\label{lvector0}
\end{align}
${\cal L}_0^{(V)}$ depends on the background Weyl multiplet,
and $(\cdots)|_{\rm conf}$ in (\ref{wel})
represents the substitution of the
conformally flat background.
In particular, the $Sp(1)_R$ gauge field $V_\mu^a$
vanishes in (\ref{wel}).
$[\mbox{CS}]_5^{ijk}$ is the 5d Chern-Simons term
defined by
\begin{equation}
[\mbox{CS}]_5^{ijk}=\epsilon^{\lambda\mu\nu\rho\sigma}
A_\lambda^i\partial_\mu A_\nu^j\partial_\rho A_\sigma^k
\end{equation}
for Abelian gauge fields.
For non-Abelian gauge fields $A^3dA$ and $A^5$ terms should be appropriately supplemented.
The prepotential ${\cal F}(\phi)$
is a homogeneous cubic polynomial of the scalar fields $\phi^i$,
and ${\cal F}_i$, ${\cal F}_{ij}$, and ${\cal F}_{ijk}$ are
its derivatives:
\begin{equation}
{\cal F}_i=\frac{\partial{\cal F}}{\partial\phi^i},\quad
{\cal F}_{ij}=\frac{\partial^2{\cal F}}{\partial\phi^i\partial\phi^j},\quad
{\cal F}_{ijk}=\frac{\partial^3{\cal F}}{\partial\phi^i\partial\phi^j\partial\phi^k}.
\end{equation}
If all the vector multiplets are
not backgrounds but dynamical
the theory is conformal.
We will mention the non-conformal case later.
The second term in (\ref{wel}) is the
curvature coupling of the scalar fields.

We would like to reproduce these transformation laws and the Lagrangian
by the supergravity.
In the 5d ${\cal N}=1$ supergravity, the Killing equation
(\ref{generalkilling}) is realized if
\begin{align}
&V_\mu^a=0,
\label{vflat}\\
&v_{\mu\nu}+2f_{\mu\nu}=0.
\label{cfbkg}
\end{align}
Indeed, if these are satisfied,
the transformation law of gravitino
in (\ref{deltapsieta}) becomes
\begin{equation}
\delta_Q\psi_\mu=D_\mu\xi-\gamma_\mu(\sla f+t)\xi,
\end{equation}
and the condition $\delta_Q\psi_\mu=0$ gives the Killing equation with
\begin{equation}
\kappa=(\sla f+t)\xi.
\label{kappaso}
\end{equation}
It is easy to confirm that the
transformation laws
(\ref{deltavector}) agree with (\ref{vectortrlaws})
if we shift the auxiliary fields
by
\begin{equation}
D'^i_a=D_a^i-2\phi^i t_a.
\end{equation}
We can also show that the Lagrangian
(\ref{wel}) is
reproduced from the supergravity Lagrangian.
In the 5d ${\cal N}=1$ supergravity
vector multiplets couple to the Weyl multiplet
through the Lagrangian
((2.11) in \cite{Kugo:2000af})
\begin{equation}
e^{-1}{\cal L}_{\rm SUGRA}^{(V)}=e^{-1}{\cal L}_0^{(V)}+e^{-1}{\cal L}_1^{(V)},
\end{equation}
where ${\cal L}_0^{(V)}$ is the Lagrangian in (\ref{lvector0}),
and ${\cal L}_1^{(V)}$ is given by
\begin{align}
e^{-1}{\cal L}_1^{(V)}
&={\cal F}P
\nonumber\\
&-i{\cal F}_iF_{\mu\nu}^i(v^{\mu\nu}+2f^{\mu\nu})
+
\frac{1}{4}
{\cal F}_{ij}
\lambda^i(\sla v+2\sla f)\lambda^j
\nonumber\\
&+
(\mbox{terms with $\psi_{\mu I}$ or $\eta_I$}).
\label{lvec1}
\end{align}
$P$ in the first line is defined by
\begin{equation}
P=
C
-20t_at_a
-4 f_{\mu\nu}v^{\mu\nu}
-6 f_{\mu\nu}f^{\mu\nu}.
\end{equation}
In the background (\ref{gensol})
this is rewritten as
\begin{align}
P&=3\left(J_{\wh m\wh n}^af_{\wh m\wh n}+2t^a\right)^2
-\frac{S^2}{4}\epsilon_{\wh m\wh n\wh p\wh q}{\cal W}_{\wh m\wh n}{\cal W}_{\wh p\wh q}
\nonumber\\
&+2D^{(4)}_{\wh m}v^{\wh m\wh 5}
-\frac{4}{S}(\partial_{\wh m}S)v^{\wh m\wh 5}
-\frac{3}{S^2}(\partial_{\wh m}S)^2.
\end{align}
The two terms in the second line in (\ref{lvec1}) contain
$v^{\mu\nu}+2f^{\mu\nu}$,
and vanish if (\ref{cfbkg}) holds.
What we need to show is that the
first line in
(\ref{lvec1}) is the same as the
curvature coupling in (\ref{wel}).
This is easily shown by using
the condition $\delta_Q\eta=0$.
If (\ref{cfbkg}) holds, we can rewrite
$\delta_Q\eta$ in (\ref{deltapsieta}) as
\begin{align}
\delta_Q\eta=4[\sla D(\sla f+t)]\xi
+4\gamma_\mu(\sla f+t)\gamma^\mu(\sla f+t)\xi
+(C-20t_at_a+2f_{\mu\nu}f^{\mu\nu})\xi.
\end{align}
Using this and $D_\mu\xi=\gamma_\mu\kappa$ with $\kappa$ in
(\ref{kappaso}), we obtain
\begin{equation}
P\xi=(C-20t_at_a+2f_{\mu\nu}f^{\mu\nu})\xi
=-4D_\mu D^\mu\xi=\frac{R}{4}\xi.
\label{vbyr}
\end{equation}
The third equality is shown by using $D_\mu\xi=\gamma_\mu\kappa$ as
follows:
\begin{align}
\frac{1}{4}R\xi
&=-\frac{1}{8}\gamma^{\mu\nu}R_{\mu\nu\rho\sigma}\gamma^{\rho\sigma}\xi
\nonumber\\
&=-\gamma^{\mu\nu}D_\mu D_\nu\xi
\nonumber\\
&=
-\sla D\sla D\xi
+D_\mu D^\mu\xi
\nonumber\\
&=
-5\sla D\kappa
+D_\mu D^\mu\xi
\nonumber\\
&=
-5D_\mu D^\mu\xi
+D_\mu D^\mu\xi
\nonumber\\
&=
-4D_\mu D^\mu\xi.
\end{align}
We used the flatness of the $Sp(1)_R$ connection
between the first and the second lines.
(\ref{vbyr}) shows that the first line in
(\ref{lvec1}) is precisely the same as the curvature coupling in (\ref{wel}).

Next, let us consider hypermultiplets.
For simplicity we consider a neutral on-shell hypermultiplet
that is not coupled by vector multiplets.
A hypermultiplet consists of scalar fields ${\cal A}^I_A$
and a symplectic Majorana fermion field $\zeta_A$.%
\footnote{We use the convention in \cite{Kugo:2000hn} for hypermultiplets.}
(See Table \ref{table:hyper}.)
\begin{table}[htb]
\caption{
Component fields in a hypermultiplet.}
\label{table:hyper}
\begin{center}
\begin{tabular}{rlccc}
\hline
\hline
& fields & $Sp(1)_R$ & $Sp(1)_F$ &  \\
\hline
bosons & scalar fields & ${\bm2}$ & ${\bm2}$ & ${\cal A}_A^I$ \\
fermion & symplectic Majorana & ${\bm1}$ & ${\bm2}$ & $\zeta_A$ \\
\hline
\end{tabular}
\end{center}
\end{table}
$A=1,2$ is an $Sp(1)_F$ flavor index.
The local supersymmetry transformation laws for the hypermultiplet are
((4.4) in \cite{Kugo:2000hn})
\begin{align}
\delta_Q{\cal A}_A^I&=2(\xi^I\zeta_A),\nonumber\\
\delta_Q\zeta_A&=-(\sla D{\cal A}^I_A)\xi_I+{\cal A}_A^I(-3t\xi-\sla f\xi+\sla v\xi)_I,
\label{hyperq}
\end{align}
and the Lagrangian is
((3.1) in \cite{Kugo:2000af})
\begin{equation}
e^{-1}{\cal L}_{\rm SUGRA}^{(H)}
=e^{-1}{\cal L}_0^{(H)}
+e^{-1}{\cal L}_1^{(H)},
\end{equation}
where ${\cal L}_0^{(H)}$
and ${\cal L}_1^{(H)}$
are given by
\begin{align}
e^{-1}{\cal L}_0^{(H)}
&=D_\mu{\cal A}_I^A D^\mu{\cal A}_A^I-2(\zeta^A\sla D\zeta_A),
\nonumber\\
e^{-1}{\cal L}_1^{(H)}
&=
\left(\frac{1}{4}R-\frac{1}{4}P-\frac{1}{4}(v_{\mu\nu}+2f_{\mu\nu})^2\right){\cal A}^A_I{\cal A}_A^I
\nonumber\\
&-\frac{1}{2}(\zeta^A\gamma_{\mu\nu}\zeta_A)(v_{\mu\nu}+2f_{\mu\nu})
\nonumber\\
&+(\mbox{terms with $\psi_{\mu I}$ or $\eta_I$}).
\label{l1h}
\end{align}

By substituting (\ref{vflat}) and (\ref{cfbkg}) into
the transformation laws (\ref{hyperq})
we obtain
the superconformal transformation laws
\begin{align}
\delta_{\rm SC}{\cal A}_A^I&=2(\xi^I\zeta_A),\nonumber\\
\delta_{\rm SC}\zeta_A&=-(\sla D{\cal A}^I_A)\xi_I-3{\cal A}_A^I\kappa_I,
\label{hypersc}
\end{align}
which are obtained from those in the flat background
by the Weyl transformation.
For the Lagrangian, the Weyl transformation
gives
\begin{equation}
e^{-1}{\cal L}_{\rm SC}^{(H)}=
e^{-1}{\cal L}_0^{(H)}|_{\rm conf}+\frac{3R}{16}{\cal A}^A_I{\cal A}_A^I,
\label{hypercf}
\end{equation}
and the curvature
coupling of the scalar fields ${\cal A}_A^I$
is reproduced by substituting
(\ref{vflat}) and (\ref{cfbkg}) into ${\cal L}_1^{(H)}$
in the same way as the vector multiplets.

Notice that
the number of the solutions
to $\delta_Q\psi_\mu=0$ is at most $8$,
and the formulation with the Poincar\'e supergravity
cannot reproduce all the $16$ supersymmetries
in the 5d superconformal algebra.
This is because
the relation (\ref{kappaso}) partially breaks
the supersymmetry
in the superconformal theory.
This can be interpreted as the supersymmetry breaking by
a mass deformation.
Mass deformations in the superconformal theory can be realized by
coupling global symmetry currents to the central charge vector multiplet
\cite{Kugo:2000hn}:
a background vector multiplet with a constant scalar component.
The components of the central charge vector multiplet are
\begin{equation}
(\phi,A_\mu,\lambda,D_a)
=(1,2ia_\mu,0,0).
\label{central}
\end{equation}
If we substitute this into
$\delta_{\rm SC}\lambda=0$, we obtain
\begin{equation}
0=\delta_{\rm SC}\lambda
=2i[\kappa-(\sla f+t)\xi],
\label{sgauge}
\end{equation}
and this is nothing but the relation (\ref{kappaso}).
Even if we consider a conformal theory,
the Weyl multiplet of the Poincare supergravity
contains the central charge vector multiplet as
a submultiplet, and it breaks a part of the superconformal symmetry.

It is shown in \cite{Fujita:2001kv} that we can construct a conformal supergravity
by separating the central charge vector multiplet from the Weyl multiplet.
In the context of the conformal supergravity $\kappa$ can be regarded as the parameter of the $S$-transformation.
The Poincare supergravity is reproduced from the conformal supergravity
by fixing the $S$ and $K$ symmetries.
(See Appendix D in \cite{Fujita:2001kv}.)
The $S$ symmetry is gauge fixed by setting the fermion component of
the central charge vector multiplet to be $0$, and
(\ref{sgauge}) defines the compensating $S$-transformation
necessary to keep the $S$-gauge fixing condition invariant
under the $Q$-transformation in the Poincare supergravity.

\subsection{${\bm S}^5$}\label{s5.ssec}
The supersymmetric theories on the round ${\bm S}^5$
and the corresponding supergravity background are
given in \cite{Hosomichi:2012ek}.
Let us confirm that this is a special case of the
solution (\ref{gensol}).

The ${\bm S}^5$ metric represented as the Hopf fibration over $\CC P_2$ is
\begin{equation}
ds^2=ds_{\CC P_2}^2+e^{\wh 5}e^{\wh 5},\quad
ds_{\CC P_2}^2=e^{\wh m}e^{\wh m},\quad
e^{\wh 5}=r(dx^5+{\cal V}),
\end{equation}
where $r$ is the radius of ${\bm S}^5$ and
${\cal V}$ is a one-form on $\CC P_2$.
We take a local frame such that $ J^3$ is the complex structure
of the $\CC P_2$,
and then the following relations hold.
\begin{equation}
S=r,\quad
{\cal W}=\frac{2i}{r^2}J^3.
\end{equation}
Due to the K\"ahlerity, the holonomy of $\CC P_2$ is $U(2)=Sp(1)_r\times U(1)_l$
where $U(1)_l\subset Sp(1)_l$ is the stabilizer subgroup of
the complex structure $J^3$.
The spin connection of $\CC P_2$
commutes with $ J^3$,
and takes the form
\begin{equation}
\omega_{\wh m\wh n}^{\CC P_2}=\frac{3i}{2}{\cal V} J_{\wh m\wh n}^3+(\mbox{$Sp(1)_r$ part}).
\end{equation}

Let us assume the invariance of the matter Lagrangians
${\cal L}_{\rm SUGRA}^{(V)}$ and ${\cal L}_{\rm SUGRA}^{(H)}$
under the $SO(6)$ rotational symmetry of the ${\bm S}^5$.
The second line of ${\cal L}_1^{(V)}$ in (\ref{lvec1})
and the second line of ${\cal L}_1^{(H)}$ in (\ref{l1h})
depend on the tensor fields $f_{\mu\nu}$ and $v_{\mu\nu}$
through the combination
\begin{equation}
v'_{\mu\nu}=v_{\mu\nu}+2f_{\mu\nu}.
\end{equation}
The $SO(6)$ invariance requires $v'^{\mu\nu}=0$,
and the independent fields should satisfy
\begin{equation}
v_{\wh m\wh 5}=0,\quad
f_{\wh m\wh n}J_{\wh m\wh n}^a+2t^a=-\frac{i}{r}\delta^{a3}.
\label{indeps5}
\end{equation}
The components of the $Sp(1)_R$ gauge field are
\begin{equation}
V_{\wh m}^a=-\frac{3i}{2}{\cal V}_{\wh m}\delta^a_3,\quad
V_{\wh 5}^a=\frac{3i}{2r}\delta^a_3.
\end{equation}
This is a flat connection and can be gauged away.
Then this solution becomes a special case of the
conformally flat background we considered in \ref{conf.ssec}.
Although
(\ref{indeps5})
do not completely fix the background fields
the ambiguity does not
affect the Lagrangians ${\cal L}_{\rm SUGRA}^{(V)}$ and ${\cal L}_{\rm SUGRA}^{(H)}$,
and they are given by
(\ref{wel}) and (\ref{hypercf}) with $R=20/r^2$.

For a mass deformed theory the Lagrangian depends
on the tensor field $f_{\mu\nu}$ through the central charge vector multiplet
(\ref{central}).
Then the $SO(6)$ invariance requires $f_{\mu\nu}=0$,
and (\ref{indeps5}) is replaced by the stronger conditions
\begin{equation}
v_{\wh m\wh 5}=0,\quad
f_{\wh m\wh n}=0,\quad
t^a=-\frac{i}{2r}\delta^{a3}.
\label{indeps52}
\end{equation}
This agree with the background fields given in \cite{Hosomichi:2012ek}.

Although a superconformal theory on the round ${\bm S}^5$
has $16$ supersymmetries,
as we mentioned in \ref{conf.ssec},
the supergravity formulation reproduces only a part of them.
For the background specified by (\ref{indeps52})
$\delta_Q\eta=0$ is automatically holds and $\delta_Q\psi_\mu=0$ gives
\begin{equation}
D_\mu\xi=-\frac{i}{2r}\tau_3\gamma_\mu\xi.
\end{equation}
This has eight solutions belonging to the
real representation
$({\bm4},{\bm2})+(\ol{\bm4},{\bm2})$
of $SO(6)\times Sp(1)_R$.

If we choose another background satisfying
(\ref{indeps5}) we obtain a different Killing spinor equation.
Although different backgrounds
give the same superconformal Lagrangians
${\cal L}^{(V)}_{\rm SC}$ and
${\cal L}^{(H)}_{\rm SC}$,
the number of supersymmetries which are realized
by the supergravity in general depends on the choice of the
background fields.

\subsection{${\bm S}^4\times\RR$}
A supersymmetric theory on ${\bm S}^4\times\RR$ can be easily
obtained by using Weyl rescaling from the theory on the
flat background,
and is used in \cite{Kim:2012gu} for the computation of
the superconformal index.
Although we can easily construct a supersymmetric background
with the geometry ${\bm S}^4\times\RR$ by using the solution
(\ref{gensol})
it gives a theory different from the Weyl-rescaled one.

Let us identify $\RR$ with the fifth direction.
$S$, $e_m^{\wh n}$, and ${\cal V}_m$ are given by
\begin{equation}
S=(\mbox{positive constant}),\quad
e_m^{\wh n}=(\mbox{vielbein of round ${\bm S}^4$}),\quad
{\cal V}_{\wh m}=0.
\end{equation}
We assume the $SO(5)$ rotational invariance of the
Lagrangians of vector and hypermultiplets.
As in the case of ${\bm S}^5$,
this requires $v_{\mu\nu}'\equiv v_{\mu\nu}+2f_{\mu\nu}=0$
for a conformal theory and $v_{\mu\nu}=f_{\mu\nu}=0$ for
a mass deformed theory.
For independent fields
these are rewritten as
\begin{equation}
v^{\wh m\wh 5}=f_{\wh m\wh n}J_{\wh m\wh n}^a+2t^a=0,
\end{equation}
for the conformal case and
\begin{equation}
v^{\wh m\wh 5}=f_{\wh m\wh n}=t^a=0,
\end{equation}
for the mass deformed case.
The latter background is given in \cite{Pan:2013uoa}.
In both cases
\begin{equation}
P=0
\label{pzero}
\end{equation}
and the $Sp(1)_R$ connection is the instanton configuration
related to the spin connection on ${\bm S}^4$ by
\begin{equation}
V^a=\frac{1}{4}\omega^{({\bm S}^4)}_{\wh p\wh q} J_{\wh p\wh q}^a.
\label{vons4}
\end{equation}
(\ref{pzero}) and (\ref{vons4})
are different from what are expected
in a Weyl-rescaled theory:
$P=R/4=3/r^2$ and flat $V_\mu^a$.
Actually it is impossible to realize a flat $Sp(1)_R$ connection
in the solution (\ref{gensol})
because ${\bm S}^4$ does not admit an almost complex structure.
It is necessary to turn on a non-trivial $Sp(1)_R$ flux for the
existence of $J_{\wh m\wh n}^a$.

This result does not change even if we take a different $x^5$ direction.
Because an arbitrary rotation of ${\bm S}^4$ has fixed points
and $R^\mu$ is nowhere vanishing,  we cannot take $x^5$
within ${\bm S}^4$ and $R^\mu$ necessarily has the component along $\RR$.
Then the topology of the base manifold $\cal B$ is ${\bm S}^4$,
and the existence of $J_{\wh m\wh n}^a$ requires non-trivial $Sp(1)_R$ flux.
Therefore, we cannot realize the Weyl-rescaled theory on ${\bm S}^4\times\RR$
as a special case of the background (\ref{gensol}).

\subsection{${\bm S}^3\times \Sigma$}\label{s3sigma.sec}
The last example we consider is
${\bm S}^3\times \Sigma$,
the direct product of three-sphere ${\bm S}^3$ with radius $r$
and a Riemann surface $\Sigma$.
A supersymmetric theory on this background
is constructed in \cite{Kawano:2012up} for $\Sigma=\RR^2$
and in \cite{Fukuda:2012jr} for general $\Sigma$.
It can be reproduced by the solution (\ref{gensol})
as is shown below.

We treat ${\bm S}^3$ as the Hopf fibration over ${\bm S}^2$,
and identify the Hopf fiber direction with $x^5$.
The metric of ${\bm S}^3\times \Sigma$ is
\begin{align}
ds^2=ds_\Sigma^2+ds_{{\bm S}^2}^2+e^{\wh 5}e^{\wh 5},\quad
ds_\Sigma^2=e^{\wh 1}e^{\wh 1}+e^{\wh 2}e^{\wh 2},\quad
ds_{{\bm S}^2}^2=e^{\wh 3}e^{\wh 3}+e^{\wh 4}e^{\wh 4},\quad
e^{\wh 5}=r(dx^5+{\cal V}),
\end{align}
where $\cal V$ is a one-form on ${\bm S}^2$.
The following equations hold.
\begin{align}
S=r,\quad
\omega_{\wh3\wh4}^{{\bm S}^2}=2{\cal V},\quad
{\cal W}=\frac{2}{r^2}e^{\wh 3}\wedge e^{\wh 4}.
\end{align}
We can take a local frame such that
$J^3$ is the complex structure of ${\bm S}^2\times\Sigma$,
which is the summation of
the complex structures of ${\bm S}^2$ and $\Sigma$.

Let us assume that the Lagrangians
${\cal L}^{(V)}_{\rm SUGRA}$
and ${\cal L}^{(H)}_{\rm SUGRA}$
are invariant under the $SO(4)$ isometry
of ${\bm S}^3$.
As in previous subsections,
all components of $v'_{\mu\nu}$ should vanish except for $v'_{\wh1\wh2}$
for the $SO(4)$ invariance.
This requires that the independent fields satisfy
\begin{equation}
v^{\wh m\wh 5}=f_{\wh m\wh n}J^a_{\wh m\wh n}+2t^a=0,
\label{vfjt}
\end{equation}
and then the non-vanishing component of $v'_{\mu\nu}$ is
\begin{equation}
v'_{\wh 1\wh 2}=\frac{1}{r}.
\end{equation}
The $Sp(1)_R$ connection is
\begin{equation}
V^a_{\wh m=\wh 1,\wh 2}=-\frac{i}{2}\delta^{a3}\omega^{(\Sigma)}_{\wh m\wh1\wh2},\quad
V^a_{\wh m=\wh 3,\wh 4}=i\delta^{a3}{\cal V}_{\wh m},\quad
V^a_{\wh 5}=-\frac{i}{r}\delta^{a3}.
\label{sp1rons3sigma}
\end{equation}
The ${\bm S}^3$ part of the connection (\ref{sp1rons3sigma})
\begin{equation}
V^{({\bm S}^3)a}=V^a_{\wh3}e^{\wh3}+V^a_{\wh4}e^{\wh4}+V^a_{\wh5}e^{\wh5}
=-i\delta^a_3dx^5
\end{equation}
is flat, and can be gauged away.
This guarantees
the $SO(4)$ invariance of ${\cal L}_0^{(V)}$ and ${\cal L}_0^{(H)}$.
The $Sp(1)_R$ connection on $\Sigma$
is topologically twisted
in such a way that a covariantly constant spinor on $\Sigma$ exists.

If the conditions in (\ref{vfjt}) are satisfied,
${\cal L}_1^{(V)}$ and ${\cal L}_1^{(H)}$ are given by
\begin{align}
e^{-1}{\cal L}_1^{(V)}
&=-\frac{2i}{r}{\cal F}_iF_{\wh1\wh2}^i
+\frac{1}{4r}{\cal F}_{ij}(\lambda^i\gamma_{\wh1\wh2}\lambda^j),
\nonumber\\
e^{-1}{\cal L}_1^{(H)}
&=
\frac{1}{r^2}{\cal A}^A_I{\cal A}_A^I
-\frac{1}{r}(\zeta^A\gamma_{\wh1\wh2}\zeta_A).
\end{align}
Although (\ref{vfjt}) does not completely determine the background fields,
the ambiguity does not affect the Lagrangians
in the absence of mass deformations
with the central charge vector multiplet.
The hypermultiplet Lagrangian ${\cal L}_{\rm SUGRA}^{(H)}$
for this background
agrees with the Lagrangian in \cite{Fukuda:2012jr}
up to field redefinition.

In the mass-deformed case
for the $SO(4)$ invariance
only non-vanishing component of $f_{\mu\nu}$
should be $f_{\wh1\wh2}$
which is related to $t^3$ by
\begin{equation}
t^3=if_{\wh1\wh2}.
\end{equation}
If we take the prepotential
${\cal F}=(1/2g_{\rm YM}^2)\phi^0\tr(\phi)^2$ with $\phi^0=1$ being the scalar component
of the central charge vector multiplet,
${\cal L}_0^{(V)}$ and ${\cal L}_1^{(V)}$ are given by
\begin{align}
e^{-1}{\cal L}_0^{(V)}
&=\frac{1}{g_{\rm YM}^2}\tr\bigg[\frac{1}{4}F_{\wh\mu\wh\nu}^2+\frac{1}{2}(D_{\wh\mu}\phi)^2
-\frac{1}{2}D_a'^2
-\frac{1}{2}(\lambda\sla D\lambda)
+\frac{1}{2}(\lambda[\phi,\lambda])
\nonumber\\
&
+
f_{\wh1\wh2}
\left(
2i\phi F_{\wh1\wh2}
+2i\phi D_3'
-\frac{1}{2}(\lambda\gamma_{\wh1\wh2}\lambda)
-\frac{i}{2}(\lambda\tau_3\lambda)
-[\mbox{CS}]_3
\right)\bigg],
\nonumber\\
e^{-1}{\cal L}_1^{(V)}
&=\frac{1}{g_{\rm YM}^2}\tr\left[
-\frac{2i}{r}\phi F_{\wh1\wh2}
+\frac{1}{4r}(\lambda\gamma_{\wh1\wh2}\lambda)
+\frac{2}{r}f_{\wh1\wh2}\phi^2\right],
\label{family}
\end{align}
where
$[\mbox{CS}]_3$ is the Chern-Simons term on ${\bm S}^3$
\begin{equation}
[\mbox{CS}]_3=\epsilon^{\wh1\wh2\mu\nu\rho}
\left(A_\mu\partial_\nu A_\rho-\frac{2i}{3}A_\mu A_\nu A_\rho\right).
\end{equation}
(\ref{family}) gives a family of the supersymmetric Yang-Mills Lagrangian
parameterized by $f_{\wh1\wh2}$,
which is a function on $\Sigma$.
For the gauge invariance of the Chern-Simons term,
the $U(1)_Z$ flux on $\Sigma$ should be quantized as
\begin{equation}
\frac{1}{g_{\rm YM}^2}\int_\Sigma f\in\frac{i}{4\pi}\ZZ.
\end{equation}
The supersymmetric Yang-Mills Lagrangian in
\cite{Fukuda:2012jr} is obtained up to
a field redefinition
by setting
\begin{equation}
f_{\wh1\wh2}=-it^3=\frac{1}{2r}.
\end{equation}

\section{Discussion}\label{disc.sec}
We constructed supersymmetric backgrounds of a 5d ${\cal N}=1$ supergravity.
We solved the supersymmetry conditions $\delta_Q\psi_\mu=\delta_Q\eta=0$,
and obtained the solution that depends on
the independent fields
\begin{equation}
S(x^m),\quad
{\cal V}_m(x^m),\quad
e_m^{\wh n}(x^m),\quad
a_m(x^m),\quad
v_{\wh m\wh 5}(x^m),
\end{equation}
on which no local constraints are imposed.
A supersymmetric background is specified by choosing these functions.
We also showed that the independent fields in the solution
can be freely changed by combining
$Q$-exact deformations and gauge transformations.
This means that the partition function does not affected
by the local degrees of freedom.

We should emphasize that we did not take care about global issues.
In order to determine the parameter dependence
of the partition function,
we need to investigate global obstructions carefully.
For example, for a compact background manifold,
we cannot freely change the fifth component of a gauge field
by gauge transformations and it may affect the partition function.
Similarly, if the manifold has non-trivial two-cycles
we have the restriction that
a flux through the cycles should be appropriately quantized.
This prohibit continuous deformations of background gauge fields,
and may cause background dependence of the partition function.
Detailed analysis of these restrictions is necessary to understand
parameter dependence of the partition function.
We hope we could return to this problem in near future.

Important feature of the solution is the existence of the isometry.
This suggests a close relation to four-dimensional supersymmetric backgrounds.
It would be interesting to study supersymmetric configurations of
4d ${\cal N}=2$ off-shell supergravity \cite{deWit1980,deWit1981}
and their relation to the solution obtained in this paper.

In Section \ref{examples.sec} we reproduced some
known examples as special cases of the general solution.
We also found that our solution
does not include all the known supersymmetric backgrounds.
A possible reason for this is
that
we assumed for simplicity that the supersymmetry parameter $\xi$
satisfies the symplectic Majorana condition.
Another possibility is that the choice of the supergravity
is not suitable to realize some of supersymmetric backgrounds.

Our analysis was based on
a Poincar\'e supergravity.
As is mentioned in \ref{conf.ssec} we cannot reproduce
all supersymmetries of a superconformal theory
in the framework of Poincar\'e supergravity.
To realize a superconformal theory
it would be more suitable to use a conformal supergravity
to describe curved backgrounds.
As is shown in \cite{Fujita:2001kv}, the Weyl multiplet shown in
Table \ref{table:weyl}
is obtained by fixing a part of the local superconformal symmetry
by using a vector multiplet as a compensator.
It is also possible to
write down the gauge fixing condition
by using
a hypermultiplet \cite{Bergshoeff:2004kh}
or
a linear multiplet \cite{Coomans:2012cf,Ozkan:2013nwa} instead of a vector multiplet.
It may be possible
to obtain a more general class of solutions
by considering a system consisting of
a superconformal Weyl multiplet
and different kinds of matter multiplets
without gauge fixing conditions imposed.

\section*{Acknowledgments}
Y.~I. is partially supported by Grant-in-Aid for Scientific Research
(C) (No.24540260), Ministry of Education, Science and Culture, Japan.
H.~M. acknowledges the financial support from the
Center of Excellence Program by MEXT, Japan through the
``Nanoscience and Quantum Physics'' Project of the Tokyo
Institute of Technology.

\appendix
\section{Appendix}\label{app.sec}
We use Greek letters $\mu,\nu,\ldots=1,\ldots,5$ for 5d world indices,
and hatted Greek letters $\wh\mu,\wh\nu,\ldots=\wh 1,\ldots,\wh 5$ for
orthonormal indices.
Roman letters $m,n,\ldots$ and $\wh m,\wh n,\ldots$ are vector
indices running over $1,\ldots,4$ or $\wh 1,\ldots,\wh4$.

The 5d anti-symmetric tensor $\epsilon^{\mu\nu\rho\sigma\tau}$
is defined by
\begin{equation}
\gamma^{\mu\nu\rho\sigma\tau}
=\epsilon^{\mu\nu\rho\sigma\tau}{\bm1}_4.
\end{equation}

We use $\alpha,\beta,\ldots=1,2,3,4$ for $Sp(2)_L$ spinor indices
and $I,J,\ldots=1,2$ for $Sp(1)_R$ doublet indices.
They are raised and lowered by $Sp(2)_L$ and $Sp(1)_R$ invariant
anti-symmetric tensors $\epsilon_{IJ}=\epsilon^{IJ}$ and $C_{\alpha\beta}=C^{\alpha\beta}$ satisfying
\begin{equation}
\epsilon^{IK}\epsilon_{JK}=\delta^I_J,\quad
C^{\alpha\gamma}C_{\beta\gamma}=\delta^\alpha_\beta.
\end{equation}

We use NW-SE convention for implicit contraction of these indices.
For example,
$(\eta\chi)\equiv \eta^{\alpha I}\chi_{\alpha I}
\equiv C^{\alpha\beta}\epsilon^{IJ}\eta_{\beta J}\chi_{\alpha I}$.

For a rank $n$ anti-symmetric tensor $A_{\mu_1\cdots\mu_n}$
we define
\begin{equation}
\sla A=\frac{1}{n!}A_{\mu_1\cdots\mu_n}\gamma^{\mu_1\cdots\mu_n}.
\end{equation}

For $Sp(1)_R$ triplet fields
we use the matrix notation
\begin{equation}
t_I{}^J\equiv t_a(\tau_a)_I{}^J
\end{equation}
where $\tau_a$ ($a=1,2,3$) are the Pauli matrices.
As an example, we present $\delta_Q\eta$ in (\ref{deltapsieta})
with all indices explicit;
\begin{align}
\delta_Q\eta_{I\alpha}&=
-2(\gamma_\nu)_\alpha{}^\beta\xi_{I\beta} D_\mu v^{\mu\nu}
+\xi_{I\alpha} C
+4(D_\mu t^a)(\gamma^\mu)_\alpha{}^\beta(\tau_a)_I{}^J\xi_{J\beta}
\nonumber\\&
+8\left(\frac{1}{2}f_{\mu\nu}(\gamma^{\mu\nu})_\alpha{}^\beta
-\frac{1}{2}v_{\mu\nu}(\gamma^{\mu\nu})_\alpha{}^\beta\right)
t^a(\tau_a)_I{}^J\xi_{J\beta}+(\gamma^{\mu\nu\rho\sigma})_\alpha{}^\beta\xi_{I\beta} f_{\mu\nu}f_{\rho\sigma}.
\end{align}

We use a convention in which a
symplectic Majorana spinor $\chi_\alpha{}^I$ is expressed
in the form
\begin{equation}
\chi=(\chi_\alpha{}^I)
=\begin{pmatrix}
U \\
D
\end{pmatrix},
\quad
U=U_0{\bm1}_2+iU_a\tau_a,\quad
D=D_0{\bm1}_2+iD_a\tau_a,
\end{equation}
with real
$U_i$ and $D_i$ ($i=0,1,2,3$),
and the scalar product of two symplectic Majorana spinors
are given by
\begin{equation}
(\chi^{(1)}\chi^{(2)})=
2U_i^{(1)}U_i^{(2)}
+2D_i^{(1)}D_i^{(2)}.
\end{equation}
Therefore,
$(\chi\chi)>0$ for a non-vanishing
Grassmann-even symplectic Majorana spinor $\chi$.
The following formulas for Grassmann-even spinors $\eta$ and $\chi$ are useful.
\begin{equation}
(\eta\chi)=(\chi\eta),\quad
(\eta\gamma_\mu\chi)=(\chi\gamma_\mu\eta),\quad
(\eta\tau_a\chi)=-(\chi\tau_a\eta).
\label{etachi}
\end{equation}
For Grassmann-odd spinors, the signs in
(\ref{etachi}) are flipped.

We do not rely on a particular choice of $\gamma^{\wh m}$, $C$, and $\epsilon$
except in \ref{s3sigma.sec}, where
we use the following matrices
\begin{equation}
\gamma^{\wh 1,\wh2,\wh3}=\begin{pmatrix}  & -i\tau_{1,2,3} \\ i\tau_{1,2,3} &  \end{pmatrix},\quad
\gamma^{\wh 4}=\begin{pmatrix}  & {\bm1}_2 \\ {\bm1}_2 &  \end{pmatrix},\quad
\gamma^{\wh 5}=\begin{pmatrix} {\bm1}_2 &  \\  & -{\bm1}_2 \end{pmatrix},
\end{equation}
\begin{equation}
\epsilon_{12}=
\epsilon^{12}=+1,\quad
C_{\alpha\beta}
=C^{\alpha\beta}
=\begin{pmatrix} \epsilon & 0 \\
0 & \epsilon \end{pmatrix}.
\end{equation}
With this choice of the matrices,
$\epsilon_{\wh\mu\wh\nu\wh\rho\wh\sigma\wh\tau}$ and $J_{\wh m\wh n}^a$
have the components
\begin{align}
&\epsilon_{\wh1\wh2\wh3\wh4\wh5}=+1,\nonumber\\
&J^a_{\wh b\wh c}=-i\epsilon_{abc},\quad
J^a_{\wh b\wh4}=i\delta^a_b\quad
(a,b,c=1,2,3).
\end{align}

\end{document}